\documentclass[twocolumn]{aastex631}

\begin{document}

\title{Studying the Radiation of a White Dwarf Star Falling onto a Black Hole}

\correspondingauthor{Marek Niko{\l}ajuk, Tomasz Karpiuk}
\email{m.nikolajuk@uwb.edu.pl, t.karpiuk@uwb.edu.pl}

\author[0000-0003-4075-6745]{Marek Niko{\l}ajuk}
\affiliation{Faculty of Physics, University of Bia{\l}ystok, ul. K. Cio{\l}kowskiego 1L, 15-245 Bia{\l}ystok, Poland}

\author[0000-0001-7194-324X]{Tomasz Karpiuk}
\affiliation{Faculty of Physics, University of Bia{\l}ystok, ul. K. Cio{\l}kowskiego 1L, 15-245 Bia{\l}ystok, Poland}

\author[0000-0002-9989-538X]{Lorenzo Ducci}
\affiliation{Institut f\"ur Astronomie und Astrophysik, Kepler Center for Astro and Particle Physics, Eberhard Karls Universit\"at, Sand 1, 72076 T\"ubingen, Germany} 
\affiliation{Department of Astronomy, University of Geneva, Chemin d'Ecogia 16, CH-1290 Versoix, Switzerland}

\author[0000-0002-5273-0641]{Miros{\l}aw Brewczyk}
\affiliation{Faculty of Physics, University of Bia{\l}ystok, ul. K. Cio{\l}kowskiego 1L, 15-245 Bia{\l}ystok, Poland}

\begin{abstract}

We investigate electromagnetic and gravitational radiation generated during the process of the
tidal stripping of a white dwarf star circulating a black hole. 
We go beyond Chandrasekhar's ideas and not only consider a white dwarf itself as a quantum object, 
but also describe the dynamics of a produced accretion disk  in a quantum way.
We model the white dwarf star  as a Bose-Fermi droplet and use the quantum hydrodynamic equations 
to simulate the evolution of the black hole-white dwarf binary system. While going through the periastron, the white dwarf 
loses a small fraction of its mass. The mass falling onto a black hole is a source of 
powerful electromagnetic and gravitational radiation. Bursts of ultraluminous radiation are 
flared at each periastron passage. This resembles the recurrent flaring of X-ray sources 
discovered recently by Irwin {\it et al.}. Gravitational energy bursts occur mainly through 
emission at very low frequencies. The accretion disk, formed due to stripping of the white 
dwarf, starts at some point to contribute continuously to radiation of both electromagnetic 
and gravitational types.
\end{abstract}


\section{Introduction} \label{sec:intro}

Black hole (BH)-white dwarf (WD) binaries are challenging to detect. However, given the known population 
of WDs, the predicted population of BHs, and several possible formation channels, 
such systems are expected to exist \citep{Ergma98,Tutuk07}. One place to search for them is among 
ultracompact X-ray binaries (UCXBs), a subgroup of low-mass X-ray binaries with very short orbital 
periods ($< 80$ minutes) and a degenerate donor star, which can be a white dwarf transferring material 
to a compact object (either a neutron star or a BH; see the recent catalog of \citealt{ArmasPadilla23}). 
So far, two UCXBs have been identified as candidates hosting a stellar-mass BH and a white dwarf:
 a source in the globular cluster 47~Tuc \citep{MillerJones15} and IGR~J17285-2922 \citep{Stoop21}.
 
Interestingly, there are also sources in external galaxies that may be explained by peculiar 
WD-BH systems. In particular, a search of X-ray data from the nearby galaxies 
NGC~4636 and NGC~5128 has uncovered two sources of ultraluminous X-ray flares 
not associated with galactic centers \citep{Irwin16}.
All these X-ray bursts have similar rise time, which is less than 1 minute 
and a decay time of about an hour. The first source flared once, with an estimated peak 
X-ray luminosity of $9 \times 10^{40}$erg s$^{-1}$, while the second flared five times and it 
was $\sim 10$ times weaker. Prior to and after the flare, a 0.3-10 keV luminosity of the 
first source was of $7.9 \times10^{38}$ erg s$^{-1}$ and $4 \times10^{37}$ erg s$^{-1}$,
respectively, in the case of the second source.

\cite{Sivakoff05} reported on another similar X-ray source in the 
elliptical galaxy NGC 4697 which showed two flares, and more recently, 
\cite{Tiengo22} reported a recurrent flaring activity from the ultra-luminous X-ray 
source XMMU J122939.7+075333 in a globular cluster in the galaxy NGC 4472 
(see \citealt{Dage24} for further discussion).
The properties of these sources are difficult to reconcile with those of better known 
Galactic accreting X-ray sources, such as the type I and II bursts observed in 
some low-mass X-ray binaries and beamed emission from stellar mass black holes 
(\citealt{Irwin16}; see, however, \citealt{Sivakoff05,Maccarone05}). Therefore, it has 
been proposed that they might constitute a new type of fast transients \citep{Irwin16}.
An explanation of observed X-ray bursts as originating from a tidal stripping of a white 
dwarf (WD) circulating an intermediate-mass black hole has been 
proposed \citep{Krolik11,Shen19,Karpiuk21,Tiengo22}.
Recently two other papers \citep{Miniutti19,Arcodia21} reported data showing sequences of 
quasi-periodic highly energetic X-ray eruptions, with possible explanation that they might be 
driven by the disruption of a compact object orbiting a supermassive black hole 
\citep{Arcodia21,Liu23}.


In this article we study the dynamics of a model white dwarf star in the field of a black 
hole. The tidal stripping of the WD by the BH causes radiation which is at the centre of 
our attention. We propose to model a cold white dwarf star by the Bose-Fermi droplet 
consisting of ultracold bosonic and fermionic atoms. Such systems have been recently 
considered theoretically \citep{Rakshit18,Rakshit18a,Karpiuk20} and should be possibly 
realized experimentally. An atomic Bose-Fermi droplet exists because some forces driving 
system's components counteract with each other. The one is attraction between bosons and 
fermions. When it is strong enough then the fermion-mediated interactions between bosons make 
them to be effectively attractive and the whole system becomes unstable due to collapse. The 
counteracting force is related to quantum pressure of fermionic component. It acts against 
the collapse of atomic Bose-Fermi droplet just like the pressure of degenerate electrons 
prohibits the gravitational collapse and stabilize, to some extent, a white dwarf star. The 
white dwarfs, in fact, very slowly evolve towards an ultimate equilibrium state of the black 
dwarf. Similarly, the Bose-Fermi droplets are finally stabilized by the action of higher 
order quantum corrections to the interatomic forces \citep{Rakshit18,Rakshit18a}.

An idea of using cold atomic gases as quantum simulators of systems otherwise inaccessible is 
already well established and sounds in accordance with an old proposition given by
\cite{Feynman82,Feynman86}. For example, this idea has been applied to study a quantum phase 
transition from a superfluid to a Mott insulator by using a gas of ultracold rubidium atoms 
in an optical lattice \citep{Greiner02} and to investigate critical phenomena near continuous 
phase transitions \citep{Hung11}. A two-dimensional caesium atoms' superfluid was quenched to 
higher interparticle interactions to observe the Sakharov oscillations, which were predicted 
to occur at the inflation era of the early Universe and are responsible for detected 
fluctuating cosmic microwave background \citep{Hung13}. Recently, the Hawking temperature of 
an analogue black hole was measured experimentally in agreement with the Hawking's theory 
\citep{Steinhauer19}. An analogue fluid black hole was created in the Bose-Einstein 
condensate of rubidium atoms. 

The surface temperatures of white dwarf stars, while forming, are of the order of $10^5\,$K 
\citep{Goldsbury12}. Their interior temperatures are much higher, of the order of $10^7\,$K 
\citep{Mestel52,ST83}. They cool down and the surface temperature reaches $2\times 10^4\,$K 
after hundred million years. This temperature falls to $10^4\,$K over the next billion years 
\citep{Fontaine01}. 
The oldest observed white dwarfs have temperatures of the order of $10^3\,$K 
\citep{Kaplan14,Kepler16}. 

High densities of WDs ($\lesssim 10^7$ g cm$^{-3}$) lead to very high Fermi temperatures ($
\sim 10^{10}$ K). These temperatures are much larger than the initial interior temperatures 
of WDs ($\sim 10^7$ K, \citealt{Mestel52,ST83}). For this reason, electrons can be treated as 
a gas at zero temperature. For bosonic component, for a WD density of $10^7$ g cm$^{-3}$ the 
critical temperature for Bose-Einstein condensation of $^4$He nuclei is $10^6\,$K. The 
assumption that the bosonic component becomes condensed, while the white dwarf cools down, 
seems to be plausible. One could argue against such a scenario based on considerations 
implying that below a certain value of temperature, the white dwarfs should crystallize 
\citep{ST83}. However, as first discussed in \cite{Gabadadze08b}, the crystallization is 
expected to occur only for white dwarfs composed of carbon, oxygen, and heavier elements. 
But, due to quantum effects (zero-point oscillations), helium white dwarfs may not solidify. 
For helium white dwarfs the melting temperature is smaller than the critical temperature for 
bosonic condensation, therefor there exists a temperature window when charged $^4$He nuclei 
can form a charged condensate \citep{Gabadadze08a,Gabadadze08b,Gabadadze09,Mosquera10}. We are 
following this line of reasoning and model a white dwarf star as a Bose-Fermi droplet. 
A great deal of knowledge that is devoted to white dwarf stars can be found in the following review articles \citep{Fontaine08,Winget08,Althaus10,Corsico19,Saumon22}.

In this work, just for simplicity, we consider the bosonic and fermionic components of the WD star 
as gases which have zero initial temperature. 
This temperature is free and can change during the simulations. 
Since we study the tidal stripping of a WD circulating a stellar-mass BH, some 
violent dynamics is expected and the falling matter might get heated. Within our description 
of bosons, with the so-called classical field approximation (for review see \citealt{review}), 
which allows for simultaneous treatment of thermal and condensed particles we can address the 
question how hot the falling matter is. 

Modeling the disruption of a WD by a BH, assuming the white dwarf star is represented by a 
cosmic Bose-Fermi droplet, was first considered by us in \cite{Karpiuk21}. This work 
improves on the previous one in some respects: now we study emission of electromagnetic and 
gravitational waves from the WD-BH binary system from the first principles and follow the 
formation of the accretion disk from the radiation point of view. 

We would like to emphasize that our modeling of the tidal disruption events of a white dwarf is 
based on quantum mechanics. We go beyond the original ideas of Chandrasekhar and not only consider 
the white dwarf itself as a quantum object, but also describe its dynamics in a quantum way. Such an attempt is 
justified and necessary because cold helium WDs are intrinsically quantum systems and the expectation of the 
appearance of qualitatively new (quantum) phenomena is justified.
Then our modeling represents yet another approach 
to tidal disruption of WDs, complementing already existing reach achievement in the field of 
dynamics of WDs (for a review of 'smoothed particle hydrodynamics', grid based, and other 
descriptions see review articles, for example, \citealt{Rosswog15,Lodato20}).

\section{Radiation from a black hole-white dwarf binary system}

Oscillating electric charge radiates electromagnetic waves \citep{Hertz}, in accordance with 
Maxwell's theory \citep{Maxwell}. Localized charge density can be expanded into multipoles 
at any time. Since the total charge of a binary system is zero, the lowest order 
contribution to radiation comes from the oscillating electric dipole moment. The power 
radiated by a localized but otherwise arbitrary electric dipole moment $\bf{p}$, calculated 
in the far-field approximation \citep{Jackson,Griffiths}, is
\begin{eqnarray}
P = \frac{2}{3\, c^3}\, |\ddot{{\bf p}}|^2  \,,
\label{totalpower}
\end{eqnarray}
where CGS units are used, $c$ is the speed of light, and double dots symbol over the electric dipole moment means the second derivative in time. In the above formula the electric dipole moment $\bf{p}$ is given by
\begin{eqnarray}
&&{\bf{p}} = \int  {\bf{r}}\, \rho_{el} ({\bf{r}})\,  d^3 r   \,,
\label{dipole}
\end{eqnarray}
where the charge density is $\rho_{el} ({\bf{r}}) = q_B n_B ({\bf{r}}) - q_F n_F ({\bf{r}})$, while $n_B ({\bf{r}})$ ($n_F ({\bf{r}})$) is the particle number density for bosons (fermions) and $q_B$ and $q_F$ are the electric charges of bosonic and fermionic components' elements building a white dwarf star. Since a white dwarf star is electrically neutral, i.e., $q_B N_B = q_F N_F$ ($N_B$ and $N_F$ are total numbers of bosonic and fermionic particles, respectively), we have
\begin{eqnarray}
{\bf{p}} &=& q_B N_B  \int  {\bf{r}}\,  \Big( \bar{n}_B ({\bf{r}}) - \bar{n}_F ({\bf{r}}) \Big)\,  d^3 r  \nonumber  \\
&=& q_B N_B \, \bar{{\bf{p}}}  \,,
\label{dipole1}
\end{eqnarray}
where 
\begin{eqnarray}
\bar{{\bf{p}}} = \int {\bf{r}}\, \Big( \bar{n}_B ({\bf{r}}) - \bar{n}_F ({\bf{r}}) \Big)\,  d^3 r 
\label{dipole2}
\end{eqnarray}
and particle number densities $\bar{n}_B ({\bf{r}})$ and $\bar{n}_F ({\bf{r}})$ are here normalized to one.

The next order contribution to electromagnetic radiation originates from a change of the electric quadrupole moment $Q_{\alpha\,\beta}$, where indices $\alpha$ and $\beta$ refer to Cartesian components (contribution coming from a time variation of the magnetic dipole moment remains of the same order) \citep{Jackson}
\begin{eqnarray}
P = \frac{1}{180\, c^5}   \sum_{\alpha,\beta}  (\dddot{Q}_{\alpha\,\beta})^2\,   \,,
\label{quadrupole}
\end{eqnarray}
where
\begin{eqnarray}
Q_{\alpha\,\beta} &=& \int  (3\, r_\alpha r_\beta - r^2 \delta_{\alpha\,\beta})\, \rho_{el} ({\bf{r}})\,  d^3 r   \nonumber  \\
&=&  q_B N_B\, \bar{Q}_{\alpha\,\beta}   \nonumber  \\
\bar{Q}_{\alpha\,\beta} &=& \int  (3\, r_\alpha r_\beta - r^2 \delta_{\alpha\,\beta})\, \Big( \bar{n}_B ({\bf{r}}) - \bar{n}_F ({\bf{r}}) \Big)\,  d^3 r \,.  \nonumber  \\
\label{quadrupole1}
\end{eqnarray}
The total power radiated by a localized oscillating charge is an incoherent sum of contributions from different multipoles \citep{Jackson}.

Falling mass generates disturbances in the gravitational field and hence becomes the source of gravitational waves travelling in space-time \citep{Abbott16}. In the lowest order, the gravitational radiation produced in this way depends on a third time derivative of a mass quadrupole moment $Q_{\alpha\,\beta}^m$, and is given by a famous Einstein's quadrupole formula \citep{Maggiore,PJAK}
\begin{eqnarray}
P = \frac{G}{45\, c^5}   \sum_{\alpha,\beta}  \langle (\dddot{Q}_{\alpha\,\beta}^m)^2 \rangle  \,,
\label{gravrad}
\end{eqnarray}
where the average $\langle \cdot \rangle$ is an average over a characteristic time associated with gravitational waves and $G$ is the gravitational constant.
The above formula has been indirectly confirmed by measuring the decay of the orbital period with time of the binary pulsar PSR 1913+16 \citep{Hulse74,Hulse75,Taylor}. In the case of a black hole-white dwarf binary system one has
\begin{eqnarray}
Q_{\alpha\,\beta}^m &=& \int  (3\, r_\alpha r_\beta - r^2 \delta_{\alpha\,\beta})\, 
\Big( m_B n_B ({\bf{r}}) + m_F n_F ({\bf{r}}) \Big)   \,  d^3 r   \nonumber \\
&=&  m_B N_B\, \bar{Q}_{\alpha\,\beta}^m   \,,
\label{quadrupolemass}
\end{eqnarray}
$m_B$ and $m_F$ being the masses of bosonic and fermionic component particles and
\begin{eqnarray}
\bar{Q}_{\alpha\,\beta}^m &=& \int  (3\, r_\alpha r_\beta - r^2 \delta_{\alpha\,\beta})\, 
\Big( \bar{n}_B ({\bf{r}}) + \frac{m_F N_F}{m_B N_B}\, \bar{n}_F ({\bf{r}}) \Big)   \,  d^3 r  \nonumber  \\
&\approx&  \int  (3\, r_\alpha r_\beta - r^2 \delta_{\alpha\,\beta})\, \bar{n}_B ({\bf{r}}) \,  d^3 r   \,.
\label{quadrupolemass1}
\end{eqnarray}
The last term in the upper row of Eq. (\ref{quadrupolemass1}) can be safely neglected since bosonic component mass dominates in the white dwarf stars as well as in the ceasium-lithium atomic Bose-Fermi droplets, assumed in our simulations. 

\section{Quantum hydrodynamic equations}

We use the formalism of quantum hydrodynamics \citep{Madelung,Frolich,Wong,MarchDeb} to 
describe Bose-Fermi mixtures. Such a treatment was already discussed many years ago by 
\citeauthor{Wheeler} in the context of the oscillations of electrons in a many-electron atom 
induced by ultraviolet and soft X-ray photons \citep{Wheeler}. They assumed the velocity field 
of electrons is irrotational. In our study, we follow the same reasoning and apply quantum 
hydrodynamic equations both for fermionic and bosonic clouds in a droplet.

The quantum hydrodynamic equations for the Bose-Fermi mixture (read \citealt{Karpiuk21}):
\begin{eqnarray}
&&\frac{\partial\, n_{F}}{\partial t} = -\nabla \cdot \left(n_F\, {\bf{v}}_{\!F}    \right)   \nonumber \\
&&m_F\frac{\partial\, {\bf{v}}_{\!F}}{\partial t} = -\nabla\left(\frac{\delta T}{\delta n_F}+\frac{m_F}{2} {\bf{v}}_{\!F}^2 + 
\frac{\delta E_{BF}}{\delta n_F}  \right) \nonumber \\
&& \frac{\partial\, n_B}{\partial t} = - \nabla \cdot \left( n_B {\bf{v}}_{\!B} \right)  \nonumber \\
&& m_B \frac{\partial\, {\bf{v}}_{\!B}}{\partial t} = - \nabla \left( \frac{\delta E_B}{\delta n_B} + \frac{m_B}{2} {\bf{v}}_{\!B}^2 + V_q + \frac{\delta E_{BF}}{\delta n_B} \right) \;,   \nonumber \\
\label{hydroFB}
\end{eqnarray}
where the fermionic and bosonic densities ($n_F({\bf r},t)$, $n_B({\bf r},t)$) and irrotational velocity fields (${\bf{v}}_{\!F}({\bf r},t)$, ${\bf{v}}_{\!B}({\bf r},t)$) are used as main variables. Different terms in the right-hand side of Eqs. (\ref{hydroFB}) (the second and the fourth ones) represent various energy components present in the system under consideration. $T$ is the intrinsic kinetic energy of an ideal Fermi gas and is calculated including lowest order gradient correction \citep{Weizsacker,Kirznits,Oliver} and $E_{BF}$ is the boson-fermion interaction energy. $E_{BF}$ includes the mean-field term and a quantum corrections due to quantum fluctuations which is necessary to stabilize the Bose-Fermi droplet \citep{Rakshit18}. The term $E_B$ describes the interaction between bosons, including the famous Lee-Huang-Yang correction \citep{LHY57}, $V_q$ is related to the bosonic quantum pressure \citep{Madelung}. Terms proportional to the square of velocities represent the energies of macroscopic flow of bosonic and fermionic fluids. Detailed formulas are given in Appendix \ref{formulas}.

The hydrodynamic equations for fermions can be put in a form of the Schr\"odinger-like equation by using the inverse Madelung transformation \citep{Dey98,Domps98,Grochowski17,Grochowski20}. This is just a mathematical transformation which introduces the single complex function instead of density field and velocity field (which represents the potential flow) used in a hydrodynamic description. Both treatments are equivalent provided the velocity field is irrotational. Hence, Eqs. (\ref{hydroFB}) can be turned into coupled Schr{\"o}dinger-like equations for a condensed Bose field $\psi_B=\sqrt{n_B} \exp{(i \phi_B)}$ (with $n_B=|\psi_B|^2$ and ${\bf{v}}_{\!B}=(\hbar/m_B) \nabla \phi_B$) and a pseudo-wavefunction for fermions $\psi_F=\sqrt{n_F} \exp{(i \phi_F)}$ (with $n_F=|\psi_F|^2$ and ${\bf{v}}_{\!F}=(\hbar/m_F) \nabla \phi_F$): $i\hbar\, \partial \psi_B / \partial t = H^{eff}_B \psi_B$ and $i\hbar\, \partial \psi_F / \partial t = H^{eff}_F \psi_F$. The effective nonlinear single-particle Hamiltonians are
\begin{eqnarray}
H^{eff}_B &=& -\frac{\hbar^2}{2 m_B}\nabla^2 
 + g_B\, |\psi_B|^2 + \frac{5}{2} C_{LHY}\, |\psi_B|^3    \nonumber \\
&+&  g_{BF}\, |\psi_F|^2 + C_{BF}\, |\psi_F|^{8/3} A(w,\alpha)   \nonumber \\
&+&  C_{BF}\,|\psi_B|^2 |\psi_F|^{8/3}\, \frac{\partial A}{\partial \alpha} \frac{\partial \alpha}{\partial n_B}   
\label{HamBF1}
\end{eqnarray}    
and  
\begin{eqnarray}  
H^{eff}_F &=& -\frac{\hbar^2}{2 m_F}\nabla^2 
+ (1 - \xi) \frac{\hbar^2}{2 m_F} \frac{\nabla^2 |\psi_F|}{|\psi_F|} \nonumber \\
&+& \frac{5}{3} \kappa_k |\psi_F|^{4/3} + g_{BF}\, |\psi_B|^2  \nonumber \\
&+& \frac{4}{3} C_{BF}\, |\psi_B|^2 |\psi_F|^{2/3} A(w,\alpha)   \nonumber \\
&+&  C_{BF}\,|\psi_B|^2 |\psi_F|^{8/3}\, \frac{\partial A}{\partial \alpha} \frac{\partial \alpha}{\partial n_F}  
\label{HamBF}
\end{eqnarray}  
with all parameters defined in Appendix \ref{formulas}. The bosonic wave function and the fermionic pseudo-wave function are normalized to the total number of particles in bosonic and fermionic components, $N_{B,F} = \int d\mathbf{r}\, |\psi_{B,F}|^2$, respectively.

Next, the Bose-Fermi droplet is placed in the field of an artificial black hole, which is assumed to be a non-rotating black hole described by the Schwarzschild space-time metric. We use further a well-working approximation, given by \cite{Paczynsky80}, to the potential energy of a test particle moving in the  Schwarzschild metric. According to this approximation, the radial part of potential energy is replaced by $V_{PN}=-GM_{BH}/(r-R_S)$, where the Schwarzschild radius is $R_S=2GM_{BH}/c^2$. The pseudo-Newtonian potential, $V_{PN} (r)$, reproduces correctly the last stable circular orbits. Since it does not depend on the angular momentum, the motion of a test particle can be approximated just as a motion in a three-dimensional space in the field of pseudo-Newtonian potential. Hence, the equations of motion for the Bose-Fermi droplet moving in the field of a fixed black hole can be put in the form
\begin{eqnarray}
&& i \hbar \frac{\partial \psi_B}{\partial t} = (H^{eff}_B + V_{PN}\,m_B)\,\psi_B  \nonumber \\
&& i \hbar \frac{\partial \psi_F}{\partial t} = (H^{eff}_F + V_{PN}\,m_F)\,\psi_F   \,.
\label{eqmWDBH}
\end{eqnarray}  

\section{Scaling the model results to the realm of astronomical objects}
\label{sec:scaling}

We solve numerically Eqs. (\ref{eqmWDBH}) and scale all quantities to the realm of 
astronomical objects. All quantities used in numerical simulations are given in the code 
units. A typical scattering length for bosonic atoms $a_B$ ($\simeq 5$ nm) represents the 
length unit. The mass is expressed in $m_B$ -- the mass of bosonic caesium atom ($= 2.21 
\times 10^{-25}$ kg). The time unit is given in $(m_B a_B^2)/\hbar$. The last one results 
from the Schr\"odinger equation, where $E \propto \hbar^2\nabla^2/(2m)$ and therefor time unit is 
$\propto \hbar/E$.  Although the atomic droplet behaviour is simulated, the results can be scaled up to 
real astronomical sources. To achieve this, let's adopt
\begin{equation}
\begin{array}{ccc}
r_{\rm astro} \  &   = & \mathcal{A} \,\, r_{\rm num} \ [a_B] \\ 
m_{\rm astro} \  &   = & \phantom{ai}\mathcal{B} \,\, m_{\rm num} \ [m_B] \\ 
t_{\rm real}  \  &   = & \phantom{aaaaaa}\mathcal{T}\, t_{\rm num} \ [m_B a_B^2/\hbar] \ , 
\label{eq:time}
\end{array}
\end{equation}
where $\mathcal{A,B,T}$ are unknown constants.
The droplet radius $r_{\rm num} \simeq 1\mu$m 
($=200 a_B$), and, since the droplet consists of $1460$ bosons and $100$ fermions 
 in our case, its mass 
$m_{\rm num} = 1460\, m_B + 100\, m_F = 3.23 \times 10^{-22}$ kg (i.e., $m_{\rm num}= 1464\, 
m_B)$. Assuming that  $r_{\rm astro}$($=\!\!r_{wd}$) and $m_{\rm astro}$($=\!\!m_{wd}$) 
are an exemplary helium white dwarf radius and mass ($r_{wd}=0.0155 R_{\odot},m_{wd}=0.4 M_{\odot}$), 
see \cite{Althaus10,Corsico19,Saumon22}, we get $\mathcal{A}\approx 10^{13}$ and 
$\mathcal{B}\approx 2\times 10^{51}$. Hence, after scaling, Eqs. (\ref{eq:time}), the mass-radius relation for 
the Bose-Fermi droplet becomes exactly as for a real white dwarf star.

To obtain the coefficient $\mathcal{T}$ we look at how the energy is scaled. 
Since the unit of energy is given by $m\, l^2 t^{-2}$, where $m$, $l$, and $t$ are the units of mass, length, and time, respectively,
we have to multiply the numerical energy by $\mathcal{B A}^2 \mathcal{T}^{-2}$ to get the astronomical one. 
Therefore, due to Eqs. (\ref{eq:time}), one has 
\begin{eqnarray}
E_{\rm astro} \  = \frac{\mathcal{B A}^2}{\mathcal{T}^2} \, E_{\rm num} \ [{\rm \hbar^2/(m_B a_B^2)}] \ ,
\label{eq:E}
\end{eqnarray}
Continuing the scaling procedure based solely on the properties of a white dwarf star we have
\begin{eqnarray}
E_{\rm astro}^{WD} &=& \frac{\mathcal{B} \mathcal{A}^2}{\mathcal{T}^2}\, \widetilde{E_{\rm num}^{BF}}\,\, \frac{\hbar^2}{m_B a_B^2}  \,,
\label{eq:EWD}
\end{eqnarray}
where the tilde symbol means the value of the indicated quantity in the code units (which is ${\rm \hbar^2/(m_B a_B^2)}$). 
Eq. (\ref{eq:EWD}) gives the value of the coefficient $\mathcal{T}$ provided the white dwarf 
($E_{\rm astro}^{WD}$) and the Bose-Fermi droplet ($E_{\rm num}^{BF}$) total energies are 
known.
The total energy of astronomical white dwarfs ranges from $\sim -10^{44}$J (near the 
Chandrasekhar limit) to $\sim -10^{42}$J (for extremely low mass stars). The total energy of the Bose-Fermi droplet used in 
the simulations is of the order of $-1$ in code units. Then Eq. (\ref{eq:EWD}) 
yields $ \mathcal{T}\approx 9.5\times 10^3$ (for $E_{\rm astro}^{WD}=-6\times 10^{42}$J).

The scaling parameters can also be obtained from the dynamical stripping process of the white dwarf orbiting 
a black hole, which gives a better insight into the tidal disruption event.
Simulations show that the size of the accretion disk is about $200\, a_B$ which is $1/4$ of the distance 
from periastron to the black hole equal to $4\, r_{wd}$ (see Sec. \ref{NumResults}). 
Since the radius of the white dwarf star is roughly its Roche lobe at periastron 
(see equation (1) in \citealt{Shen19}) one can write
\begin{eqnarray}
\frac{2\, r_{wd}}{4} \left(\frac{M_{BH}}{m_{wd}}\right)^{1/3} &=& \mathcal{A}\,\, 200\,\, a_B  \,,
\label{adsize}
\end{eqnarray}
which gives $\mathcal{A}\approx 1.1\times 10^{13}$ 
(for $M_{\rm BH} = 5\, M_{\odot}$, see the following subsection).
The value of the outer radius of the accretion disk is calculated in the Appendix \ref{3rdMethod}, it is
$r_{\rm out} \lesssim 1.4\, r_{wd}$, which is very close to the numerical value of $r_{wd}$. 
This result indicates that although the atoms in the Bose-Fermi droplet 
interact via the van der Waals interactions, the dynamics of the droplet can mimic gravitational forces well. 
This finding highlights the relevance of our computer simulations. 

Additionally, accordingly to \cite{Shen19} (see their equation (3)),
the duration of the stripping is roughly given by the internal dynamical time scale 
$t_{\rm dyn} \simeq (r_{wd}^3/(G m_{wd}))^{1/2}$. Then the value of the scaling parameter 
$\mathcal{T}$ comes from
\begin{eqnarray}
\left(\frac{r_{wd}^3}{G m_{wd}}\right)^{1/2} &=& \mathcal{T}\,\, 10^4\,\, \frac{m_B a_B^2}{\hbar}  \,,
\label{stripping}
\end{eqnarray}
where $10^4$ is the numerical value of the stripping time (see the next section), and equals $\mathcal{T} \approx 9300$.
\subsection{Black hole mass}

Scaling, Eq. (\ref{eq:E}), has strong implications when applied to the gravitational energy of 
a binary system, leading to
\begin{eqnarray}
G\, \frac{M_{BH}\, m_{\rm astro}}{r_{\rm astro}}  &=&  \frac{\mathcal{B} \mathcal{A}^2}
{\mathcal{T}^2}\, \frac{\widetilde{(G M_{BH})}_{\rm num}\, m_{\rm num}\, \frac{\hbar^2}{m_B^2 a_B}}
{r_{\rm num}} \,,  \nonumber \\ 
\label{eq:Egrav}
\end{eqnarray}
where $\widetilde{(G M_{BH})}_{\rm num}$ is the numerical value of the product $G M_{BH}$ which enters the 
pseudo-Newtonian potential. In numerics, the mass of the black hole, $M_{BH}$, is encoded in 
the expression $G M_{BH}$. However, Eq. (\ref{eq:Egrav}) makes possible to retrieve $M_{BH}$ 
via
\begin{eqnarray}
M_{BH} &=& \frac{\mathcal{A}^3}{\mathcal{T}^2}\, \frac{\widetilde{(G M_{BH})}_{\rm num}\,\, \frac{\hbar^2}
{m_B^2 a_B}}{G}  \,.
\label{eq:MBH}
\end{eqnarray}
Hence, Eq. (\ref{eq:MBH}) allows to find the mass of the black hole which enters the binary 
system. For $\widetilde{(G M_{BH})}_{\rm num} \approx 1$, used in simulations, and 
$\mathcal{T}\approx 9.5\times 10^3$  calculated above,
Eq. (\ref{eq:MBH}) gives the mass of the black hole $M_{\rm BH} \simeq 5\, M_{\odot}$. 
Binary black hole-white dwarf systems with low-mass black holes $M_{\rm BH} \lesssim 10\, M_{\odot}$ have been already studied in \cite{Ergma98}.

\subsection{Scaling the electromagnetic radiation}

The dipolar energy radiated during the single event of a tidal stripping is (see Eqs. (\ref{totalpower}) and (\ref{dipole1}))
\begin{eqnarray}
\int P_{\rm dip}\, dt =\frac{2}{3\, c^3}\, (q_B N_B)_{\rm astro}^2\, \left( \int \ddot{\bar{{\bf p}}}^{\,2}\,  dt 
\right)_{\rm astro}  \,,
\label{powerperpulse}
\end{eqnarray}
where integration is performed over the time of the stripping. 
Since the unit of $\ddot{{\bf p}}^{\,2}$ is given
by $m\, l^5 t^{-6}$, the radiated energy, after Eq. (\ref{dipole1}), scales as
\begin{eqnarray}
\frac{2}{3\, c^3}\,  \frac{\mathcal{B}\, \mathcal{A}^5}{\mathcal{T}^5}\,  (q_B N_B)_{\rm num}
^2\,   \left( \int \ddot{\bar{{\bf p}}}^{\,2}\,  dt \right)_{\rm num}  \,.
\label{pulsescaling}
\end{eqnarray}
The efficiency of conversion of the rest-mass energy of the accreted matter into radiation 
for Schwarzschild black holes is less than $\eta \approx 0.06$ \citep{Longair}. 
Since the mass stripped off the white dwarf and falling down to the black hole during 
a single periastron 
passage is $5\times 10^{-4} m_{wd}$ (see the next section), the effective charge of bosons is limited by
\begin{eqnarray}
(q_B N_B)_{\rm num} < \left(\frac{\eta\, (5\times 10^{-4} m_{wd})\, c^2}{\frac{2}{3\, c^3}\,  
\frac{\mathcal{B}\, \mathcal{A}^5}{\mathcal{T}^5} \left( \int \ddot{\bar{{\bf p}}}^{\,2}\,  dt 
\right)_{\rm num}}\right)^{1/2} \, .
\label{BFcharge}
\end{eqnarray}
Therefore the dipolar radiation power is bounded by
\begin{eqnarray}
P_{\rm dip} < \frac{2}{3\, c^3}\,  \frac{\mathcal{B}\, \mathcal{A}^5}{\mathcal{T}^6}\,  {\rm RHS}\,   \left( 
\ddot{\bar{{\bf p}}}^{\,2}\, \right)_{\rm num}  \,,
\label{dippowerscaling}
\end{eqnarray}
where ${\rm RHS}$ is the square of the right hand side of Eq. (\ref{BFcharge}). Since
$\int \ddot{\bar{{\bf p}}}^{\,2}\, dt\,|_{num}$ within the first drop (see Fig.~\ref{sumratio10}, upper frame) is $\sim 
10^{-12}$ $\hbar^3/(m_B^3 a_B^4)$  and $\ddot{\bar{{\bf p}}}^{\,2}|_{num} \sim 10^{-15}$ 
$\hbar^4/(m_B^4 a_B^6)$ at maxima, the peak dipolar radiation power is estimated as 
$P_{\rm dip}^{\rm peak} \sim 10^{49}\,$erg/s.

On the other hand, the luminosity of the flares of electromagnetic radiation originating 
from the variation of the electric quadrupole moment, Eq. (\ref{quadrupole}), reads
\begin{eqnarray}
P_{\rm quad} < \frac{1}{180\, c^5}\,  \frac{\mathcal{B}\, \mathcal{A}^7}{\mathcal{T}^8}\,  {\rm RHS}\,\,  
\Big( \sum_{\alpha,\beta}  (\dddot{\bar{Q}}_{\alpha\,\beta})^2 \Big)_{\rm num}   \,.
\label{quapowerscaling}
\end{eqnarray}
This is because the unit of $\dddot{{Q}}_{\alpha\,\beta}^{2}$ is $m\, l^7 t^{-8}$. 
Since $\sum_{\alpha,\beta} (\dddot{\bar{Q}}_{\alpha\,\beta})^2|_{num} \sim 10^{-14}$ 
$\hbar^6/(m_B^6 a_B^8)$ at maxima, the peak quadrupole radiation power for first drops is 
estimated as $P_{\rm quad}^{\rm peak} \sim 10^{47}\,$erg/s.

\subsection{Scaling the gravitational radiation}

The power of gravitational radiation is given by Eq. (\ref{gravrad}). According to Eq. (\ref{quadrupolemass}) this radiation scales as
\begin{eqnarray}
P_{\rm grav} &=& \frac{G}{45\, c^5}\,   \Big( \sum_{\alpha,\beta}  (\dddot{Q}^m_{\alpha\,\beta})^2 \Big)_{\rm astro}   \nonumber \\
&=& \frac{G}{45\, c^5}\, (m_B N_B)_{\rm astro}^2\, \Big( \sum_{\alpha,\beta}  (\dddot{\bar{Q}}^m_{\alpha\,\beta})^2 \Big)_{\rm astro}   \nonumber \\
&=& \frac{G}{45\, c^5}\, m_{wd}^2\,\, \frac{\mathcal{A}^4}{\mathcal{T}^6} \, \Big( \sum_{\alpha,\beta}  (\dddot{\bar{Q}}^m_{\alpha\,\beta})^2 \Big|_{num}  [\hbar^6/(m_B^6 a_B^8)] \Big) \,.  \nonumber \\
\label{scalinggrav}
\end{eqnarray}
The scaling factor in the last line in Eq. (\ref{scalinggrav}) appears because the unit of 
$(\dddot{\bar{Q}}^m_{\alpha\,\beta})^{2}$ is given by ${l^4 t^{-6}}$.
Since $\sum_{\alpha,\beta} (\dddot{\bar{Q}}^m_{\alpha\,\beta})^2 |_{num} \sim 10^{-16}$ $\hbar^6/(m_B^6 a_B^8)$ at maxima (see the next section), the peak gravitational radiation power is estimated as $P_{\rm grav}^{\rm peak} \sim 10^{35}\,$erg/s.

\section{Numerical results and discussion}
\label{NumResults}

\begin{figure}[t]
\includegraphics[width=8.0cm]{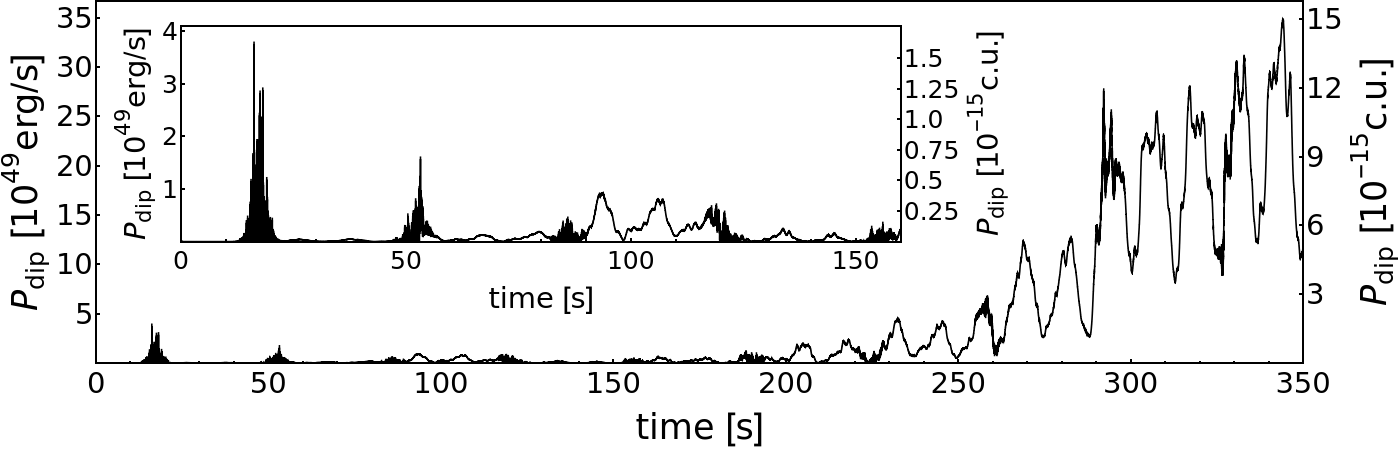}  \\   \vspace{0.4cm}
\includegraphics[width=8.0cm]{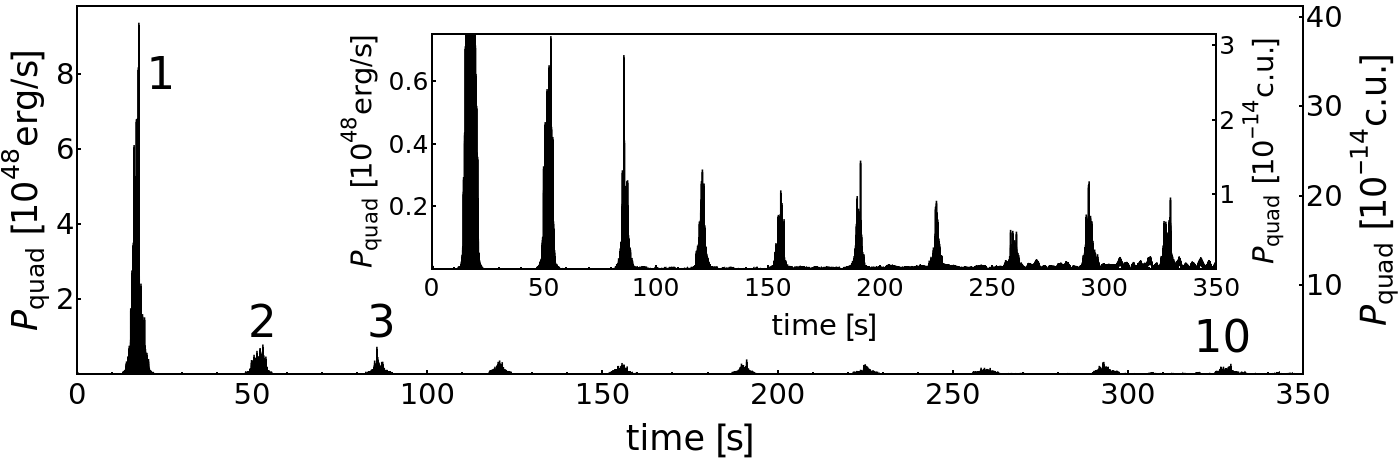}  \\   \vspace{0.4cm}
\includegraphics[width=8.0cm]{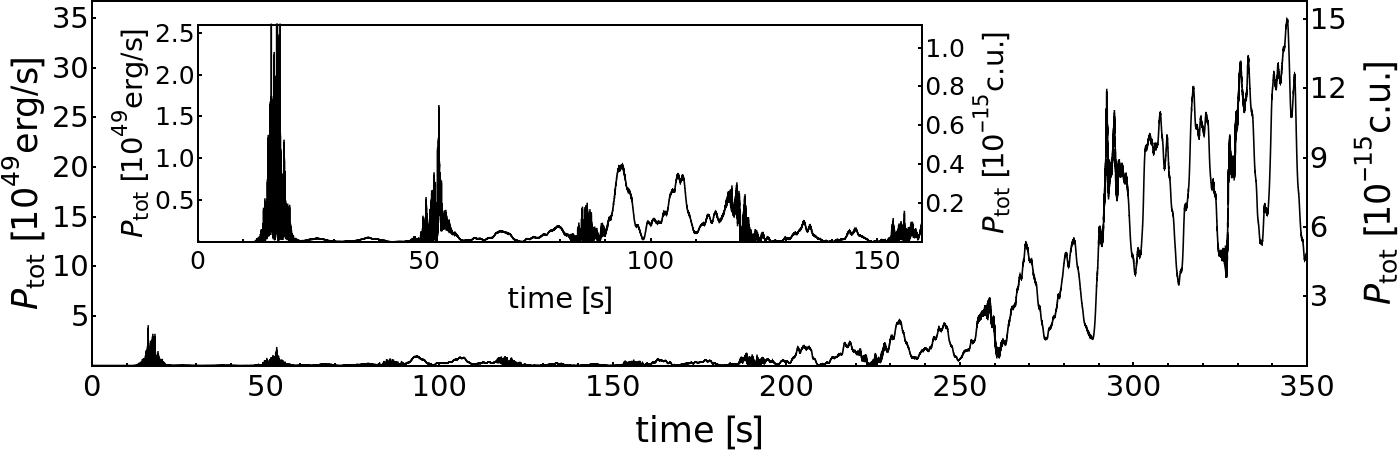}
\caption{Electromagnetic radiation power of the system of a black hole and a white dwarf as a function of time. 
Successive frames (from top to bottom) show: electric dipole contribution, electric quadrupole part of radiation 
(with some passages through periastron marked by numbers, i.e. by $1$, $2$, $3$, and $10$), and sum of 
electric dipole, magnetic dipole, and electric quadrupole contributions for the case when 
$M_{BH} = 5 M_{\odot}, m_{wd}=0.4 M_{\odot}$. }
\label{sumratio10}
\end{figure}

We solve numerically Eqs. (\ref{eqmWDBH}) by split-operator technique \citep{Gawryluk17} for 
trajectories of an atomic white dwarf corresponding to closed eccentric orbits. We consider 
the Bose-Fermi droplet consisted of $^{133}$Cs bosonic and $^{6}$Li fermionic atoms. Such 
mixtures are studied experimentally nowadays \citep{Weidemuller14,Chin17,Chin18}. However, it 
is not really important which particular atoms are chosen since the mass and the radius of the 
Bose-Fermi droplet are finally scaled to the mass and size of a real white dwarf star and the 
total energy of the droplet is of the same order as that of the astronomical WDs. 
The droplet, consisted of $1460$ bosonic and $100$ fermionic atoms, is initially located at 
apastron at some distance (about $7\, r_{\rm wd}$) from the artificial black hole, far away 
from the horizon. The initial numerical parameters are chosen in such a way that we can follow 
several revolutions of a white dwarf in an eccentric orbit (with periastron at about $4\, 
r_{\rm wd}$). Such a task is numerically manageable (i.e., is not too intensive with respect 
to the computational time) provided the white dwarf is not located extremely far from the 
black hole. This, on the other hand, means that the period of revolution of a white dwarf 
around a black hole remains short, in fact it is of the order of ten seconds 
(see Fig.~\ref{sumratio10} for $M_{BH}=5\, M_{\odot}$).
At each periastron passage a white dwarf is stripped off an approximately (at least within 
initial few passes) the same mass, which is about $5\times 10^{-4}$ of the initial white dwarf mass 
for a particular orbit discussed below. The dimensionless penetration parameter $\beta\, 
(\equiv r_t / r_{per})$, where $r_t$ is the tidal radius and $r_{per}$ is the position of the 
periastron, equals $0.73$ in the case of $M_{BH}=5\, M_{\odot}$.

Accelerating localized charged mass is a source of electromagnetic radiation. The lowest order 
contribution to radiation of a localized charge distribution, which remains neutral, comes 
from a time-dependent electric dipole moment, Eq. (\ref{totalpower}). We split the electric 
dipole moment into parts representing the accretion disk and the matter pulled out of the 
white dwarf and the white dwarf itself. In our further considerations, we focus on the 
radiation coming from the accretion disk and the falling matter.

In Fig.~\ref{sumratio10} we summarize our results for the case of stellar-mass black holes. We 
show the electromagnetic radiation power of a binary system (with $M_{\rm BH}=5\, M_{\odot}$)
as a function of time, restricting ourselves to two lowest multipoles. We also include the 
frame (the bottom one) which depicts the incoherent sum of electric dipole and quadrupole 
contributions. Actually, we also added the contribution from the magnetic dipole term, 
assuming it is equal to that related to the electric quadrupole one (this assumption seems to 
be reasonable since both terms appear at the same level of multipole expansion). Ten 
periastron passages, equally separated in time and lasting each for about $10^4\,
$code units ($\approx 4.9\,$s), can be easily recognized. Clearly, bursts of radiation emerge 
while a white dwarf is coming through the periastron. Such regular emergence of radiation 
bursts resembles recurrent flaring of X-ray sources reported recently 
\citep{Irwin16,Miniutti19,Arcodia21,Tiengo22}. Since the mass falling to the 
black hole is huge in our case (it is about $5\times 10^{-4}\, m_{wd}$ which is, in fact, by 
several orders of magnitude larger than that suggested in \cite{Shen19} to explain high 
peak luminosity of observed X-ray flares) the radiation power is enormous, limited by
$\sim 10^{49}$erg/s depending on the conversion efficiency. The amount of dipole radiation exceeds 
the quadrupole one by more than an order of magnitude for a stellar-mass black hole case, Fig. 
\ref{sumratio10} (first outburst, in fact, is excessively high, due to a kind of numerical 
turn-on effect). The total radiation (bottom frame in Fig. \ref{sumratio10}) understood as the 
incoherent sum of lowest order multipoles, including the magnetic dipole term, shows regular 
bursts as well.

\begin{figure}[t]
\includegraphics[width=8.0cm]{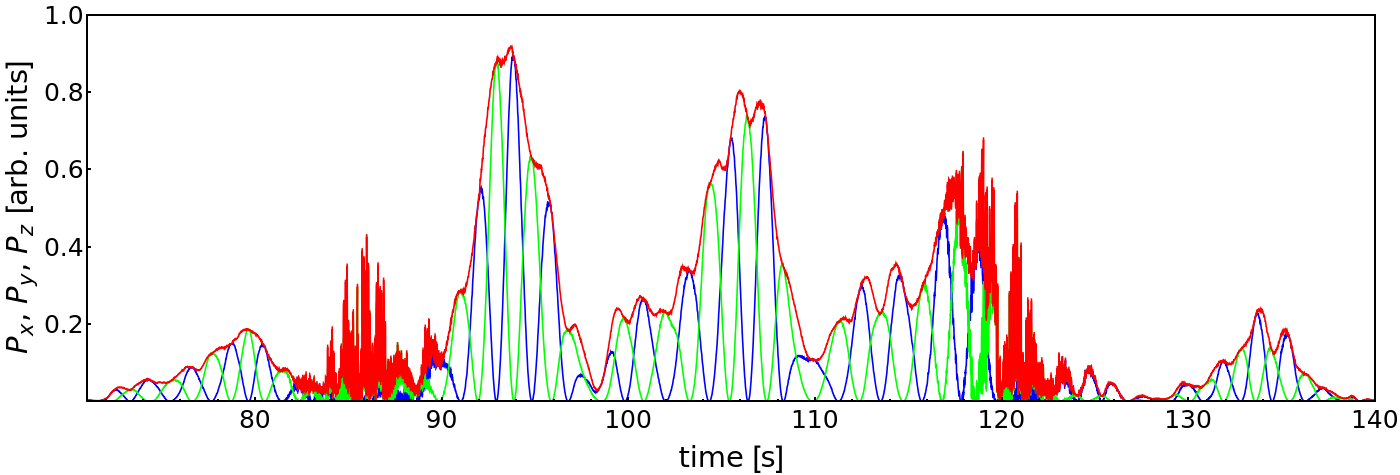} 
\caption{Power of electromagnetic radiation of the system of a black hole and a white dwarf 
emitted along $\hat{\bf{x}}$ (blue colour), $\hat{\bf{y}}$ (green colour), and $\hat{\bf{z}}$ 
(red colour) directions as a function of time, showing mainly the time period between the 
third and fourth periastron passages (see Fig. \ref{sumratio10}). Curve plotted in red colour 
is just the sum $P_x+P_y$. }
\label{powerxyz}
\end{figure}

Another interesting outcome of numerical simulations is that the accretion disk itself 
becomes a source of strong electric dipole radiation, see top frame in Fig. \ref{sumratio10}. 
Here, the accretion disk activates already after the third passage and becomes dominant at 
later time. Further properties of dipole radiation can be uncovered by looking at the dipole 
emission from a binary system in a particular direction. The power of radiation emitted in 
direction along angle $\Theta$ with respect to vector $\ddot{\bf{p}}$ is \citep{Griffiths}
\begin{eqnarray}
P \sim |\ddot{{\bf p}}|^2\,  \sin^2{\Theta}  \,.
\label{powerdir}
\end{eqnarray}
Then the radiation power detected along $\hat{\bf{x}}$, $\hat{\bf{y}}$, and $\hat{\bf{z}}$ 
directions becomes $P_{x,y,z} \sim \ddot{{\bf p}}^2\, \Big(1 - \ddot{p}_{x,y,z}^2/\ddot{{\bf 
p}}^2 \Big)$. Since $p_z=0$ one has $P_z \sim \ddot{p}^{\,2}$ and $P_x+P_y=P_z$. $P_{x,y,z}$ 
quantities are plotted in Fig. \ref{powerxyz} for $M_{BH} = 5 M_{\odot}, m_{wd}=0.4 M_{\odot}$.
There is a 
qualitative difference in the observed dipole radiation depending on the observation angle. 
Since a charged mass dropped towards an accretion disk still (although for a limited time) 
orbits a black hole, it can be considered as a rotating electric dipole. Then it can be 
thought of as a superposition of two dipoles oscillating along $x$ and $y$ axes, respectively, 
and being out of phase by $\pi/2$. The energy emitted along $x$ and $y$ axes then exhibits a 
characteristic pattern as a function of time which is $\sim (1 - \cos^2{\omega_d t})$ and $
\sim (1 - \sin^2{\omega_d t})$, respectively (see Fig. \ref{powerxyz}, blue and green curves). 
Here, $\omega_d$ is a frequency of rotating dipole (falling mass on BH) which is higher than 
the frequency of orbiting white dwarf. Therefore, between two successive periastron passages 
one observes several oscillations in $P_x$ and $P_y$, whereas no such behaviour is visible in 
time-dependence of $P_z$. The properties of dipole-type radiation can tell us about the 
orientation of the orbital plane of a white dwarf with respect to the Earth. The accretion 
disk emits much weaker quadrupole-type radiation, see Fig. \ref{sumratio10} middle frame. This 
is understandable since the structure of an accretion disk is mainly that of a rotating 
electric dipole type. 

Formation of an accretion disk is a violent process and it is expected that some part of 
falling matter becomes hot and its bosonic component can not be still condensed. This is, 
indeed, the case. The condensate in bosonic part of an accretion disk becomes depleted. We 
have estimated, by using the classical field approximation \citep[e.g.][for review]{review}, how much matter 
turns into a thermal form. The noncondensed fraction in bosonic matter in accretion disk 
becomes about $10\%$ for the revolutions below $7$th and increase up to $20\%$ for the 
revolution $9$th and successive.

\begin{figure}[t]
\includegraphics[width=8.0cm]{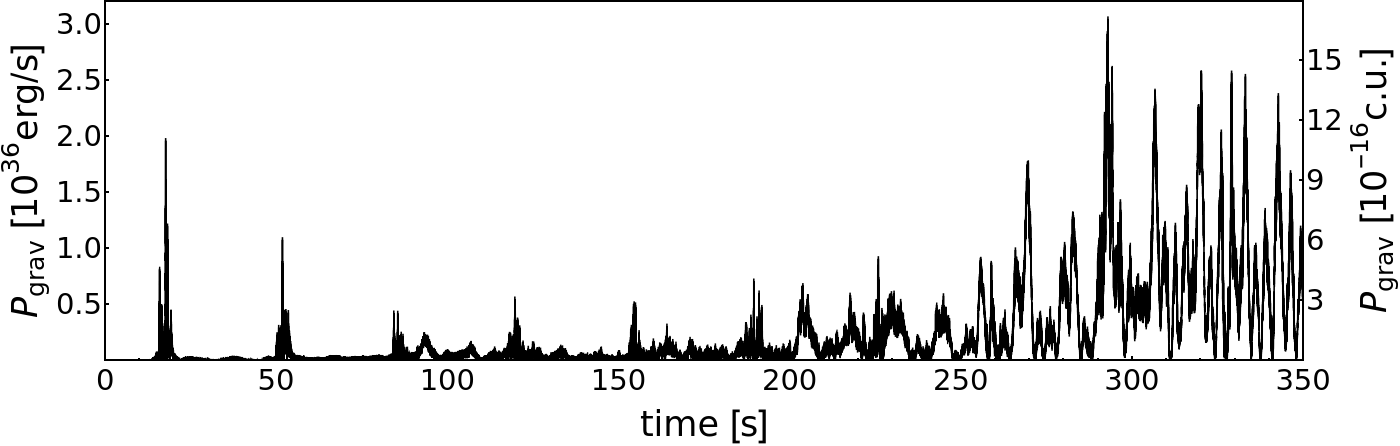}  
\caption{Gravitational radiation power of the system of a black hole and a white dwarf as a function of time for the case
$M_{BH} = 5 M_{\odot}, m_{wd}=0.4 M_{\odot}$. }
\label{sumgrav}
\end{figure}

Similar bursts of energy are visible while looking at the gravitational radiation power 
emitted by a binary black hole-white dwarf system, Fig.~\ref{sumgrav}. However, the amount of 
energy radiated via gravitational waves remains several orders of magnitudes lower than the 
energy emitted electromagnetically, compare Figs.~\ref{sumgrav} and \ref{sumratio10}. For such 
a stellar-mass black hole-white dwarf system a peak gravitational-wave luminosity at a 
periastron passage is of the order of $10^{35}\,$erg/s (see Fig.~\ref{sumgrav}, where 
instantaneous values, according to Eq. (\ref{scalinggrav}), are shown). This is, not 
surprisingly, much less than the maximum gravitational-wave luminosity detected in GW150914 
event, the first observation of gravitational waves from the merger of two stellar-mass black 
holes \citep{Abbott16}, which was $\sim 10^{56}\,$erg/s. Two years later LIGO and Virgo 
detectors made the first observation of a binary neutron star inspiral, known as the signal 
GW170817 \citep{Abbott17}. The energy radiated via gravitational waves in this event was 
estimated from below as $0.025 M_{\odot} c^2$. Combined with the knowledge of a duration of a 
gravitational wave signal, which lasted for approximately $100$ seconds, leads to average 
radiation power at least of the order of $10^{52}\,$erg/s. Merger of binary neutron stars was 
followed by a short $\gamma$-ray burst GRB 170817A \citep{Abbott17a} located in the same sky 
position, thus linking the short burst of $\gamma$ rays to coalescence of neutron stars. The 
power radiated electromagnetically by GRB 170817A event was estimated as of the order of 
$10^{46}\,$erg/s, much less than the power radiated via gravitational waves. 

\begin{figure}[h!bt]
\includegraphics[width=8.0cm]{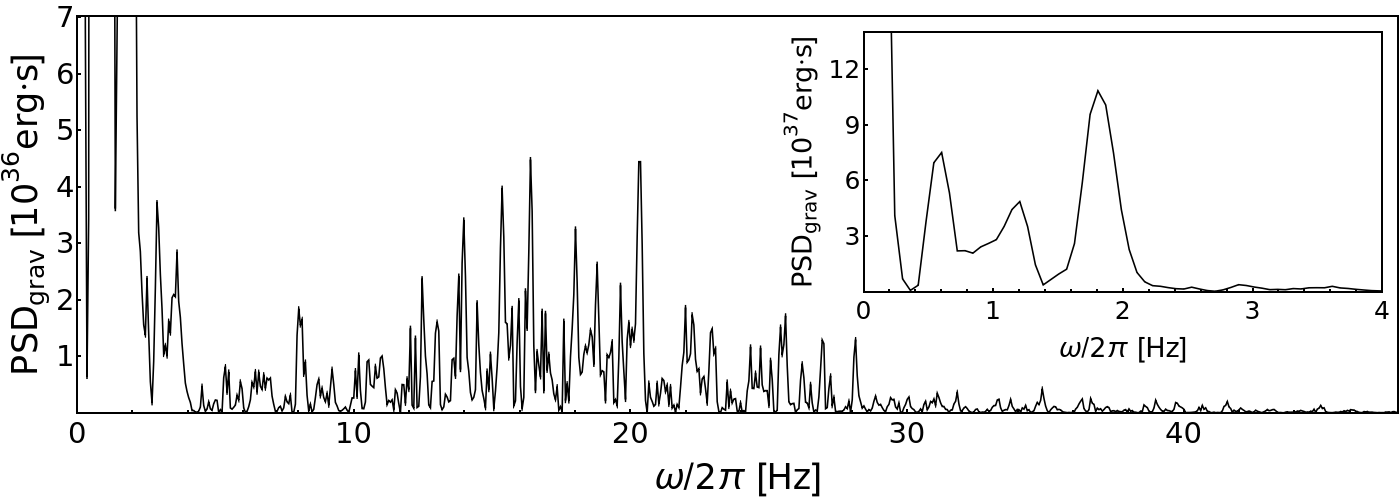}
\caption{Power spectral density of gravitational waves emitted during the second periastron 
passage (see Fig. \ref{sumgrav}) for the case $M_{BH} = 5 M_{\odot}, m_{wd}=0.4 M_{\odot}$.
Inset shows PSD for low frequencies, below $\omega/2\pi = 4\,$Hz. }
\label{omegaemitted}
\end{figure}

For the binary system studied here, with a white dwarf star moving along an eccentric orbit, 
we observe significant tidal stripping of WD accompanied by emission of electromagnetic and 
gravitational radiation. The electromagnetic bursts are much stronger than gravitational ones, 
compare Figs. \ref{sumratio10} and \ref{sumgrav}. An important question concerns the 
frequencies at which binary system emits its gravitational energy. Here, we are particularly 
focused on the second periastron passage of a white dwarf, see Fig. \ref{sumgrav}. The 
gravitational energy emitted per frequency unit, i.e. power spectral density (PSD), is given 
by the Fourier transform of gravitational radiation power, Eq. (\ref{scalinggrav}),
\begin{eqnarray}
{\rm PSD}_{grav} =  \left| {\cal{F}} \left[ \sqrt{\,{\rm P}_{grav}} \right]  \right|^2 
\label{psd}
\end{eqnarray}
and is depicted in Fig. \ref{omegaemitted}. It turns out that the energy is mainly radiated at 
low frequencies, below $1\,$Hz, see the inset (more than $95${\%} of total 
energy is emitted within the range of frequencies $\omega/2\pi < 4\,$Hz). Already planned 
gravitational-wave observatory in space, the Laser Interferometer Space Antenna (LISA), is 
supposed to work in frequency range covering the one visible in the inset in Fig. 
\ref{omegaemitted}.

\section{Conclusions}
In summary, we have studied the electromagnetic and gravitational radiation emitting from a
system of a helium white dwarf (mass about $0.4 M_{\odot}$) orbiting a stellar-mass black hole 
(with mass $\simeq 5 M_{\odot}$). For the trajectory, we study the penetration parameter 
$\beta=0.73$. As a model of the WD star, we consider a Bose-Fermi droplet of 
attractively interacting degenerate atomic bosons and spin-polarized atomic fermions, being 
initially at zero temperature. Our quantum-hydrodynamics-based simulations indeed show that 
the black hole-white dwarf binary system emits the electromagnetic energy (both in dipole and 
quadrupole modes) in bursts while crossing the periastron area, with estimated radiation 
power below $\sim 10^{49}$erg/s, depending on the radiation efficiency. 
Such regular high-energy blasts resemble the ultraluminous X-ray bursts reported by \cite{Irwin16}. 
We also find that the accretion disc, as it forms, emits energy mainly in the form of dipole rather 
than quadrupole radiation (see Fig. \ref{sumratio10}). An irregular behaviour of the radiative power 
can be seen when looking phase on to the plane of the accretion disc,
while looking edge on an oscillation pattern is present.
This property can be used to determine the orientation of the orbital plane of the 
black hole-white dwarf binary system with respect to the observer on the Earth (Fig. \ref{powerxyz}).
Our simulations demonstrate that a relative strong gravitational pulse is emitted at periastron, mainly 
at very low frequencies (Fig. \ref{omegaemitted}). Moreover, it is shown that the binary system becomes a source of 
nonmonochromatic continuous gravitational radiation as soon as the accretion disk is formed.

\begin{acknowledgments}
T.K. and M.B. acknowledge support from the (Polish) National Science Center Grant no. 2017/25/B/ST2/01943. 
This publication was created as part of the projects funded in Poland by the Minister of Science based
on agreement number 2024/WK/03.
L.D. acknowledges funding from the Deutsche Forschungsgemeinschaft (DFG, German Research Foundation) -- Projektnummer 549824807.
Part of the results were obtained using computers at the Computer Center of the University of Bialystok.
\end{acknowledgments}

\appendix

\section{Energy contributions}
\label{formulas}
\noindent
The intrinsic kinetic energy of an ideal Fermi gas is 
\begin{eqnarray}
T = \int d{\bf r}\, \left( \kappa_k\,n_F^{5/3} - \xi\, \frac{\hbar^2}{8m_F} \frac{(\nabla n_F)^2}{n_F} \right)
\label{T3D}
\end{eqnarray}
with $\kappa_k = (3/10)\,(6\pi^2)^{2/3}\,\hbar^2/m_F$ and $\xi=1/9$ 
\citep{Weizsacker,Kirznits,Oliver}.

\noindent
The boson-fermion interaction energy is
\begin{eqnarray}
E_{BF} &=& \int d{\bf r}\, g_{BF}\, n_B({\bf r}) n_F({\bf r})
+ C_{BF}\int d{\bf r}\, n_B\, n_F^{4/3} A(w,\alpha) \,,
\label{EBF}
\end{eqnarray}
where $w=m_B/m_F$ and $\alpha=2w (g_B n_B/\varepsilon_F)$ are dimensionless parameters, 
$C_{BF}=(6 \pi^2)^{2/3} \hbar^2 a_{BF}^2 / 2 m_F$, and the function $A(w,\alpha)$ is given 
in a form of integral \citep{Giorgini02} 
\begin{eqnarray}
A(w,\alpha) = \frac{2(1+w)}{3w}\left(\frac{6}{\pi}\right)^{2/3}\int^{\infty}_0 {\rm d}k \int^{+1}_{-1}{\rm d}{\Omega}
\left[ 1 -\frac{3k^2(1+w)}{\sqrt{k^2+\alpha}}
\int^{1}_0{\rm d}q q^2 \frac{1-\Theta(1-\sqrt{q^2+k^2+2kq\Omega})}{\sqrt{k^2+\alpha}+wk+2qw\Omega}  \right], \nonumber\\
\label{A}
\end{eqnarray}
with $\Theta()$ being the step theta-function.

\noindent
The bosonic quantum pressure term is 
\begin{eqnarray}
V_q=-\hbar^2/(2 m_B)\, (\nabla^2\sqrt{n_B}) /\sqrt{n_B}  \,.
\label{Vq}
\end{eqnarray}

\noindent
The interaction between bosons, including the famous Lee-Huang-Yang correction \cite{LHY57}, is
\begin{eqnarray}
E_B = g_B n_B^2/2 + C_{LHY} \int d{\bf r}\, n_B^{5/2} \ ,
\label{EB}
\end{eqnarray}
with $C_{LHY}=64/(15\sqrt{\pi})\,g_B\, a_B^{3/2}$. $g_B$ and $g_{BF}$ appearing in the above energy expressions are coupling constants for contact interactions between atoms \citep{PitaevskiiStringari}, with $g_B = 4\pi \hbar^2 a_B/m_B$ and $g_{BF} = 2\pi \hbar^2 a_{BF}/\mu$, where $a_B$ ($a_{BF}$) is the scattering length corresponding to the boson-boson (boson-fermion) interaction and $\mu=m_B\, m_F/(m_B+m_F)$ is the reduced mass.
\section{Scaling of the length}
\label{3rdMethod}
\noindent
We saw in Section \ref{sec:scaling} that $\mathcal{A}$ factor is approximately equal to
$10^{13}$, based on two methods. Let's estimate it again calculating the outer radius of the accretion disc.
In a stellar binary system, the outer radius, $r_{\rm out}$, represents the radius of the first orbit 
where the accretion disk is formed. The last stable orbit around BH is the innermost stable circular orbit (often called the ISCO).
If we assume that the particle in the Kepler orbit at $r_{\rm out}$ has the same angular momentum as it had when it passed through the Lagrange point $L_1$ \citep{FKR02}, we get 
\begin{eqnarray}
r_{\rm out} = \frac{X_{\rm L}^4}{D^3} \frac{M_{BH} + m_{wd}}{M_{BH}} \,. 
\label{eq:rcirc}
\end{eqnarray}
$D$ is the distance between BH and WD, $X_{\rm L} (\simeq  D - r_{\rm lobe})$ is the distance from the BH to the Lagrange point $L_1$ of the binary system, and $r_{\rm lobe}$ is the Roche lobe around WD. 
The radius of the WD is slightly larger than or equal to the Roche radius at periastron 
($r_{wd} \gtrsim r_{\rm lobe} \approx 0.62\,r_{wd}$, for $M_{BH} \simeq 5 M_{\odot}$ and $m_{wd}=0.4 M_{\odot}$,
see Eq. 2 in \citealt{Eggleton83}).
Thus, Eq. (\ref{eq:rcirc}) can be rewritten as
\begin{eqnarray}
r_{\rm out} = D \left(1-\frac{\tilde{r}}{D} \right)^4 \left( 1 + q \right) \ ,
\label{eq:rcirclobe}
\end{eqnarray}
with $q= m_{wd}/M_{BH}$ and $r_{lobe} < \tilde{r} < r_{wd}$.
Next, we will find the distance $D$ on the assumption that it is equal to the radius of the periastron, $r_{\rm per}$
\citep[e.g.][]{Evans15,Chen23}
\begin{eqnarray}
D \equiv r_{\rm per} = \beta^{-1} \mu \, (M_{BH}/m_{wd})^{1/3}\ r_{wd} \ ,
\label{eq:rper}
\end{eqnarray}
where $\beta$ is the dimensionless penetration parameter, and $\mu \simeq 1$.
In our case $\beta=0.73, M_{BH} \simeq 5 M_{\odot}, m_{wd}=0.4 M_{\odot}$, and $r_{wd}=0.0155 R_{\odot}$.
Thus, based on Eqs. (\ref{eq:rcirclobe})-(\ref{eq:rper}), 
we get $r_{\rm per} \simeq 3.2 \, r_{wd}$ and $0.8 \, r_{wd} < r_{\rm out} < 1.4 \, r_{wd}$.
The outer radius of numerical accretion disk is about $200\, a_B$. Thus we can write
\begin{eqnarray}
r_{\rm out} = \mathcal{A}\,\, 200\,\, a_B  \, ,
\end{eqnarray}
which gives $8.2\times 10^{12} < \mathcal{A} < 1.5\times 10^{13}$.

\bibliography{BHWDRAD}{}
\bibliographystyle{aasjournal}

\end{document}